\documentclass[%
reprint,
superscriptaddress,
amsmath,amssymb,
aps, prl,
floatfix,
]{revtex4-2}

\usepackage{graphicx}% Include figure files
\usepackage{dcolumn}% Align table columns on decimal point
\usepackage{bm}% bold math
\usepackage{color}
\usepackage{xcolor}
\usepackage{hyperref}

\begin{document}

\preprint{APS/123-QED}

\title{Reduction or enhancement of stellarator turbulence by impurities}% Force line breaks with \\
%\thanks{A footnote to the article title}%k

\author{J. M. Garc\'ia-Rega\~na}
 \email{jose.regana@ciemat.es}
\author{I. Calvo}%
\affiliation{%
 Laboratorio Nacional de Fusi\'on, CIEMAT, 28040 Madrid, Spain\\
}%
\author{F. I. Parra}
\affiliation{Princeton Plasma Physics Laboratory, Princeton, NJ 08540, USA\\
}%
\author{H. Thienpondt}%
\affiliation{%
	Laboratorio Nacional de Fusi\'on, CIEMAT, 28040 Madrid, Spain\\
}


\date{\today}% It is always \today, today,

\begin{abstract}
A systematic study of the impact of impurities on the turbulent heat fluxes is 
presented for the stellarator Wendelstein 7-X (W7-X) and, for comparison, the Large Helical Device and ITER. 
By means of nonlinear multispecies gyrokinetic simulations, it is shown that
impurities, depending on the sign of their density gradient, can significantly enhance or reduce turbulent ion heat losses.
For the relevant scenario of turbulence reduction, an optimal impurity concentration that minimizes the ion heat diffusivity emerges as a universal feature.
This result demonstrates the potential of impurities for controlling turbulence and accessing enhanced confinement regimes in fusion plasmas and, in particular, in W7-X.
\end{abstract}

\maketitle

The stellarator is a promising concept for magnetic confinement fusion reactors, 
mainly due to its intrinsic steady-state operation and absence of disruptions.
Nonetheless, the actual possibility for the stellarator to become competitive against the tokamak has historically been handicapped 
by two main difficulties: the substantially larger neoclassical particle and heat losses caused by the three-dimensionality of the magnetic field, and the more challenging engineering and manufacturing required to build stellarators, compared to tokamaks.
In relation to these two difficulties, Wendelstein 7-X (\mbox{W7-X}), the most advanced stellarator in operation, 
has represented a huge leap forward. 
First, W7-X has been optimized for reduced neoclassical transport 
 by a careful tailoring of the magnetic configuration. 
The success of this optimization strategy has been demonstrated experimentally, leading 
to record values of the fusion triple product in stellarators \cite{Beidler_nature_2021}.
 Second, despite the complexity of the 
 system of modular coils needed to produce the desired configuration, 
 measured errors are smaller than one 
 part in $100\,000$ \cite{Pedersen_nature_2016}, which meets the crucial milestone 
 of building a large optimized stellarator within restrictive tolerances.
 
W7-X has also 
brought to the forefront other questions to which less importance has traditionally been given. 
Because of the successful reduction of neoclassical losses, turbulence has become, in general, the main transport channel. 
In standard ECRH W7-X plasmas, power balance analyses show that 
heat fluxes are predominantly turbulent throughout the entire plasma volume \cite{Bozhenkov_nf_2020}. In those scenarios
a stiff turbulent heat transport has been pointed out as the reason why the core ion temperature  
clamps at approximately $1.5$ keV, which is considerably lower than the values predicted by neoclassical theory. 
So far, a higher core ion temperature has only been achieved in high performance scenarios 
with
reduced turbulence, where the neoclassical optimization of W7-X has 
been experimentally confirmed \cite{Beidler_nature_2021}. 
Such scenarios have been transiently reached by means of series of cryogenic pellet injections, which increase the density gradient of the main species and moves the operation
point from a region in the space of profile gradients where ion-temperature-gradient (ITG) turbulence is close to be maximal to a region where it is significantly attenuated. This mechanism for reducing turbulence has been confirmed by means of numerical simulations \cite{Xanthopoulos_prl_2020} and is demonstrated to be particularly efficient in W7-X, in comparison with other stellarator configurations \cite{Thienpondt_submitted_2024}. 
Since steady state operation is one of the main advantages of the stellarator concept, as well as a central goal of the W7-X project, the identification of additional mechanisms that mitigate turbulent heat losses beyond a transient phase is, therefore, pressingly required.
% In this line, based on gyrokinetic numerical simulations, the reduction of the turbulent heat flux by highly energetic particles has been proposed as another possibility \cite{DiSiena_prl_2020}.
In this Letter,	%carried out with the code \texttt{stella} \cite{Barnes_jcp_2019}},
we report on the role of impurities as a highly efficient actuator on turbulence. In order to assess the universality of the effects found,  the Large Helical Device (LHD, Japan) and ITER are also considered. We have found that impurities can 
reduce or enhance the turbulent fluctuations and associated ion heat transport. In the three devices, an optimal concentration is found that minimizes the ion heat flux. Moreover, among the three, only W7-X exhibits a remarkable reduction in electron heat transport, mirroring the trend observed for the ion heat flux with varying impurity content.
This work emphasizes the great potential of injecting impurities and controlling its concentration toward improving the plasma performance.
Moreover, it also warns about conditions where  impurities increase the turbulent heat losses.

\begin{figure}
	\includegraphics{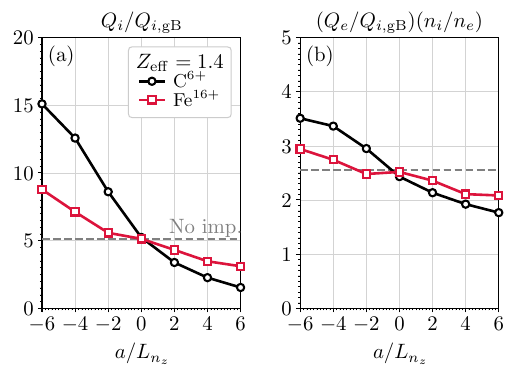}
	\vspace{-0.75cm}
	\caption{Main ion (a) and electron (b) heat fluxes as a function of the impurity density gradient in W7-X, considering Fe$^{16+}$ (squares) and C$^{6+}$ (circles), each as a single impurity in a plasma with $Z_{\mathrm{eff}}=1.4$. For all the figures of the Letter, quantities have been averaged over time during the saturated phase. In addition, heat fluxes are normalized to the ion gyro-Bohm value (see Appendix), and heat flux levels in the absence of impurities are indicated by the dashed gray line.}
	\label{fig:QiandQe}
\end{figure}

The results that follow are obtained by means of electrostatic collisionless nonlinear simulations performed in flux tube geometry with the gyrokinetic code \texttt{stella} \cite{Barnes_jcp_2019}, which has already been applied to W7-X transport studies in a number of previous works \cite{Regana_jpp_2021, Regana_nf_2021, Thienpondt_PRR_2023, Thienpondt_submitted_2024, Zocco2022, Podavini_2023}.  
All simulations are performed at the radial position \mbox{$r/a=0.7$}. Here, $r$ is the flux surface label and $a$ is the minor radius of the device. 
We assume that the plasma consists of hydrogen nuclei, electrons and a single impurity species with densities $n_{i}$, $n_{e}$ and $n_z$, respectively, all with the same temperature $T$ and normalized temperature gradient $a/L_T=3.0$. Here, the normalized gradient of a physical quantity $X$ is defined as $a/L_X=-a\,\text{d}\ln X/\text{d}r$. The equilibrium density of electrons and impurities as well as their normalized gradients are constrained by the lowest-order quasineutrality equation and its radial derivative. Namely,
\begin{eqnarray}
&n_e=n_i+Zn_z,
\label{eq:quasineutrality1}\\
&a/L_{n_e}=(n_i/n_e)\,a/L_{n_i} + (Zn_z/n_e)\,a/L_{n_z},
\label{eq:quasineutrality2}
\end{eqnarray}
with $Z$ the charge state of the impurity. Throughout the present study, the main ion density, taken as reference value, and its density gradient ($a/L_{n_i}=1.0$) have remained unchanged (see Appendix). 

\begin{figure}
	\includegraphics{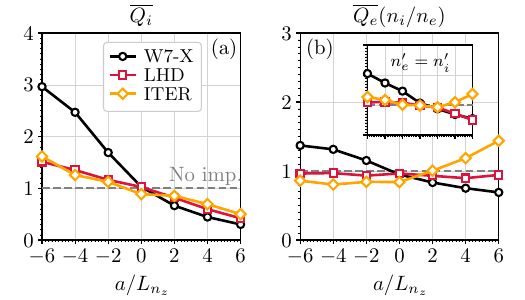}
	\vspace{-0.75cm}
	\caption{Main ion (a) and electron (b) heat fluxes as a function of the impurity density gradient, considering C$^{6+}$ and $Z_{\mathrm{eff}}=1.4$ for W7-X, LHD and ITER. The inset shows the inconsistent case where $a/L_{n_i}=a/L_{n_e}$ (see main text). Throughout this work, an overbar denotes normalization of the quantity with respect to its value in the absence of impurities.}
	\label{fig:QiandQe_others}
\end{figure}

For W7-X, we have performed a scan of the impurity density gradient, $a/L_{n_z}$, in the range $[-6, 6]$ for C$^{6+}$ and Fe$^{16+}$ impurities at concentrations such that $Z_{\textrm{eff}}=1.4$, with $Z_{\textrm{eff}}=\sum_j Z_j^2 n_j/n_e$ and $j$ running over the ion species of the plasma.
Figures ~\ref{fig:QiandQe}(a) and 	\ref{fig:QiandQe}(b) display, respectively, the ion ($Q_i$) and electron ($Q_e$) heat fluxes in gyro-Bohm units for these series of simulations. The electron heat flux is multiplied by $n_i/n_e$ in order to balance out the $Q_e$ increase due to the fact that $n_e/n_i>1$ when $Z_{\text{eff}}>1$. 
In the first place, comparing the black and red curves of, respectively, $Q_i$ and $Q_e$, with their values with no impurities, shown in gray, it can be clearly observed that both fluxes are modified by the presence of impurities. Roughly, $Q_i$ and $Q_e$ are enhanced when the impurity density is hollow and are reduced when the impurity density is peaked. Second, the impact of carbon is appreciably stronger than that of the heavier and more highly charged iron impurity.

Fig.~\ref{fig:QiandQe_others} shows, for LHD and ITER as well, $Q_i$ and $Q_e$ as a function of $a/L_{n_z}$ considering C$^{6+}$ as impurity. Although more moderate in these two devices, $Q_i$ reduction (or enhancement) occurs when $a/L_{n_z}$ is positive (or negative), as in W7-X. This dependence of $Q_i$ on $a/L_{n_z}$  strongly resembles the role that impurities have on the ITG stability in simplified geometries \cite{Tang_1980, Dominguez_1993, Dong_1994, Dong_1995}. In those references, the linear growth rate increases or decreases depending on whether the main ions and impurities have, respectively, anti-parallel or parallel density gradients. 	The results discussed so far in this Letter demonstrate that the ion heat reduction or enhancement holds for nonlinear turbulence, regardless of the magnetic geometry, and when all species are treated kinetically.
In contrast, the effect on $Q_e$ is clearly less pronounced and the trend is different from one device to another: for LHD, $Q_e$ is practically independent on $a/L_{n_z}$; for ITER, $Q_e$ increases with increasing $a/L_{n_z}$, transitioning from a slight reduction to enhancement; and only W7-X exhibits a similar behaviour for $Q_e$ as for $Q_i$.

It must be emphasized that, when impurities are included in these simulations, the only bulk plasma parameters modified are 
the electron density and its gradient, consistently with Eqs.~(\ref{eq:quasineutrality1}) and (\ref{eq:quasineutrality2}) (in the scan along $a/L_{n_z}$ from $-6$ to $6$, $a/L_{n_e}$ has been modified from $0.4$ to $1.4$).
Since $a/L_{n_e}$ is a free energy source for the electrons, one might wonder whether the differences in $Q_e$ are due to the sole presence of the impurity or due to the different impact each configuration might experience to the modification of $a/L_{n_e}$. To investigate this, we have performed simulations inconsistent with equation (\ref{eq:quasineutrality2}), forcing $a/L_{n_e}=a/L_{n_i}$, despite the presence of the impurity. The results, depicted in the inset of figure \ref{fig:QiandQe_others}(b), allow us to conclude that $Q_e$ in different configurations shows different trends mostly due to the sole presence of the impurity, and not due to $a/L_{n_e}$. This conclusion does not change qualitatively when compared with the data in the main figure. Nevertheless, a correction due to the modification of $a/L_{n_e}$, that we explore further below, can be inferred.

\begin{figure}
	\includegraphics{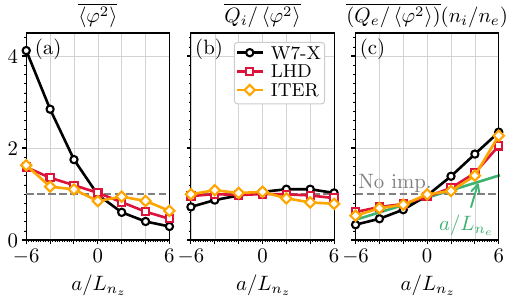}
	\vspace{-0.75cm}
	\caption{For C$^{6+}$ impurities at a concentration such that $Z_{\mathrm{eff}}=1.4$: dependence on the impurity density gradient of (a) the averaged squared potential fluctuations, $\left<\varphi^2\right>$, and (b) the main ion and (c) electron heat fluxes over $\left<\varphi^2\right>$ for W7-X, LHD and ITER. For clarity, $a/L_{n_e}$ is illustrated in (c).}
		%The electrostatic potential fluctuations are provided in units of  $T_i a/e \rho_i$  and the heat fluxes in units  of $Q_{i,\mathrm{gB}}$.}
	\label{fig:phi2_and_Q} 
\end{figure}

Beyond heat fluxes, the amplitude of the electrostatic potential fluctuations is, analogously to $Q_i$, enhanced or reduced,
% depending on whether the impurity density gradient has, respectively, the opposite or the same sign as that of the main ion density gradient
as Fig.~\ref{fig:phi2_and_Q}(a) shows for the case with C$^{6+}$ and the three analyzed devices. Figures \ref{fig:phi2_and_Q}(b) and \ref{fig:phi2_and_Q}(c) depict the ratio of, respectively, the ion and electron heat fluxes over the squared electrostatic fluctuations, $\left<\varphi^2\right>$, where $\left<...\right>$ denotes the average over the flux tube volume. For ions, $Q_i/\left<\varphi^2\right>$ depends weakly on $a/L_{n_z}$, which indicates that the ion heat flux is strongly tied to the fluctuation level of the potential. In other words, the ion heat flux decreases or increases as turbulent fluctuations do. 
In contrast, Fig.~\ref{fig:phi2_and_Q}(c) shows that $Q_e$ increases more sharply with $a/L_{n_z}$ than $\left<\varphi^2\right>$ does. Remarkably, all devices exhibit a similar trend, which roughly follows that of $a/L_{n_e}$, depicted for clarity. As $a/L_{n_e}$ is equally increased in all three devices, the alignment is not surprising. This indicates that, ultimately, despite the different impact that the presence of the impurity alone has on $Q_e$ for each device (see inset of Fig.~\ref{fig:QiandQe_others}(b)), the correction due to $a/L_{n_e}$ is comparable for all three devices, at least in this case of a moderate $Z_{\mathrm{eff}}$ of 1.4.

\begin{figure}
	\includegraphics{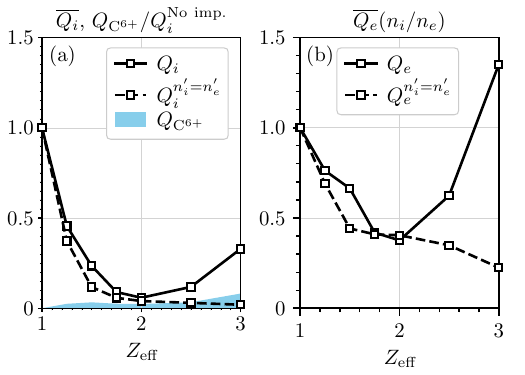}
	\vspace{-0.75cm}
	\caption{For W7-X, heat fluxes of (a) main and impurity ions, and (b) electrons as a function of $Z_{\mathrm{eff}}$, with C$^{6+}$ the only impurity present in the plasma.}
	\label{fig:Qi_and_Qe_vs_Zeff}
\end{figure}

\begin{figure}
	\includegraphics{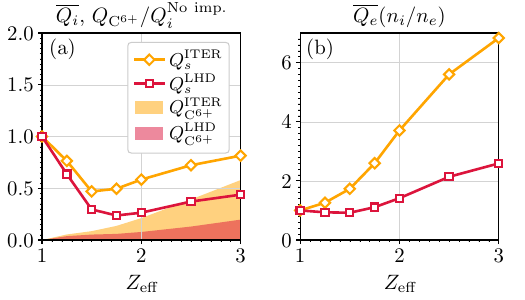}
	\vspace{-0.75cm}
	\caption{For LHD and ITER, heat fluxes of (a) main and impurity ions, and (b) electrons as a function of $Z_{\mathrm{eff}}$, with C$^{6+}$ the only impurity present in the plasma.}
	\label{fig:Qi_and_Qe_vs_Zeff_others}
\end{figure}

\begin{figure*}
	\includegraphics{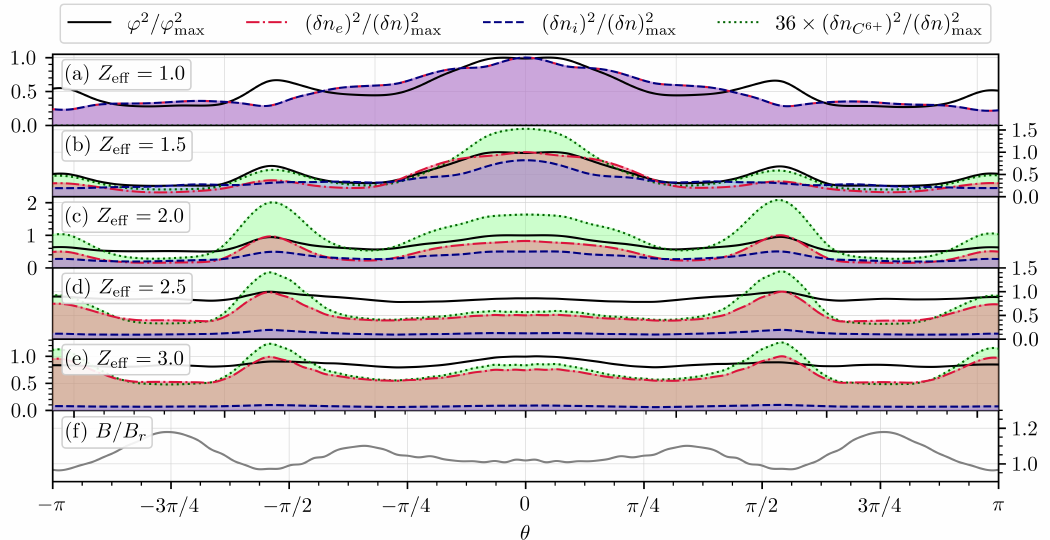}
	\caption{\label{fig:dens2_vs_z} For the $Z_{\mathrm{eff}}$ scan depicted in Fig.~\ref{fig:Qi_and_Qe_vs_Zeff}: (a)-(e) shows the normalized electrostatic potential fluctuations, $\varphi^{2}/\varphi^2_{\mathrm{max}}$, and normalized density fluctuations, $(\delta n_s)^2/(\delta n)^2_{\mathrm{max}}$, of main ions, electrons and C$^{6+}$ as a function of the poloidal coordinate; (f) shows the modulus of the magnetic field.}
\end{figure*}

%Considering the results just discussed, the deliberate injection of impurities, along with the control of the impurity content, can be viewed as a tool to access improved plasma scenarios. 
With regard to the case of turbulence reduction, it is then natural to wonder how large the impurity concentration must be in order to maximize heat flux reduction (impurity-induced reduction of turbulence in tokamak \cite{McKee2000} and stellarator \cite{Osakabe2014, Takahashi2018, Nespoli_nature_2022, Lunsford_2021} experiments is moderately reported, and simulations have occasionally pointed out to this impurity effect in tokamak geometry \cite{Pusztai_2011, Wei_2018}). To answer this question, a scan in the effective charge has been performed for the case of C$^{6+}$ with $a/L_{n_z}=6$. For W7-X, the results for the main ion and impurity heat fluxes are represented in Fig.~\ref{fig:Qi_and_Qe_vs_Zeff}(a) and for the electron heat flux in Fig.~\ref{fig:Qi_and_Qe_vs_Zeff}(b). Both $Q_i$ and $Q_e$ decrease with growing effective charge up to the value $Z_{\mathrm{eff}}=2$. The decrease is more significant for $Q_i$, which is reduced by an order of magnitude, than for $Q_e$, which is halved. In contrast, for $Z_{\mathrm{eff}}>2$ both $Q_i$ and $Q_e$ grow with $Z_\mathrm{eff}$. In the range from $Z_{\mathrm{eff}}\approx2$ to $Z_{\mathrm{eff}}\approx3$, $Q_i$ recovers approximately one third of its value in the absence of impurities, whereas $Q_e$ enhances its level clearly above that without impurities. This increasing trend is due to the increasing value of $a/L_{n_e}$ imposed by Eq.~(\ref{eq:quasineutrality2}). Ignoring Eq.~(\ref{eq:quasineutrality2}), i.e.~forcing $a/L_{n_e}=a/L_{n_i}$ (dashed lines), makes $Q_e$ and $Q_i$ decrease mononotically with $Z_{\mathrm{eff}}$ in the entire represented range.
Furthermore, the impurity heat flux consistently remains low and does not counterbalance the decrease in the heat flux of the main species. 
Similar $Z_{\mathrm{eff}}$ scans for LHD and ITER, depicted in Fig.~\ref{fig:Qi_and_Qe_vs_Zeff_others}, reveal the local minimum of $Q_i$ as a universal feature, occurring at a value of $Z_{\mathrm{eff}}\sim 1.5-2.0$. 
	On the other hand, the impurity  heat flux increases strongly with $Z_{\mathrm{eff}}$, and the electron heat flux is not decreased with increasing $Z_{\mathrm{eff}}$, in contrast to W7-X.
	However, the idea of an optimal concentration for overall reduction of turbulence holds. For instance, at $Z_{\mathrm{eff}}\sim1.75$ in LHD, $Q_i$ is minimum, and neither $Q_e$ nor $Q_{C^{6+}}$ have increased much---interestingly,
 carbon injections have traditionally assisted in accessing high-$T_i$ plasmas with internal transport barriers in LHD \cite{Takahashi2018}, and the dependence of ion heat diffusivity on carbon concentration, with an evident resemblance to our $Q_{i}(Z_{\mathrm{eff}})$ curve, has been reported in \cite{Osakabe2014}. 

Returning to the evidently interesting case of \mbox{W7-X}, the reason why $Q_e$ is not reduced as much as $Q_i$ in the range from $Z_{\mathrm{eff}}=1$ up to $2$, and why it increases more significantly than $Q_i$ beyond that range is likely due to the transition to an electron-density-gradient-driven dominated turbulence, as is shown next.
Let $(\delta n_{s})^2(\theta)$ be the squared density fluctuations of species $s$ as a function of the poloidal coordinate, and $(\delta n)^2_{\mathrm{max}}=\mathrm{max}\left\{\delta n^2_i(\theta), \delta n^2_e(\theta), \delta n^2_z(\theta), \theta \in [-\pi,\pi] \right\}$ the largest density fluctuation amplitude found \mbox{along $\theta$}. Figures \ref{fig:dens2_vs_z}(a)-6(e) show, for $Z_{\mathrm{eff}}=\left\{1.0, 1.5, 2.0, 2.5, 3.0\right\}$, $(\delta n_{s})^2(\theta)/(\delta n)^2_{\mathrm{max}}$ for the main ions, electrons and C$^{6+}$. 
%In this figure, where  $\left<\varphi^2\right>$ is also represented, one can inspect the density fluctuation distribution along the field line, the relative size of the density fluctuations among the three species and which of them dominate. 
The squared electrostatic potential fluctuations as a function of the the poloidal angle, $\varphi^2(\theta)$, normalized to its maximum value, $\varphi^2_{\mathrm{max}}=\mathrm{max}\left\{\varphi^2(\theta), \theta \in [-\pi,\pi]\right\}$, is also represented.
Examining this figure, it can be observed how the increase of $Z_{\mathrm{eff}}$ modifies the turbulent electrostatic potential, from strongly localized at the center to becoming spread out along the flux tube. Moreover, the main ion density fluctuations become significantly reduced with respect to those of the electrons, particularly for $Z_{\mathrm{eff}}=\{2.5$, $3.0\}$. 
For these high impurity concentration cases, $\delta n^2_e(\theta)$ coincides almost exactly with $Z^2\delta n^2_z(\theta)$, pointing out that the main ions do not respond to the turbulence. 
In addition, electron density fluctuations are strongly localized in the regions of magnetic field wells (see Fig.~\ref{fig:dens2_vs_z}(f) for the magnetic field strength, $B$, as a function of $\theta$). This indicates that, as $Z_{\mathrm{eff}}$ and, consequently, $a/L_{n_e}$ increase, turbulence acquires a vigorous TEM character, as we anticipated above. Such a change in the turbulence characteristics explains 
why the electron heat flux increases abruptly for $Z_{\mathrm{eff}}>2$ while the ion heat flux is not affected to the same extent in  Fig.~\ref{fig:Qi_and_Qe_vs_Zeff}. 
 
In summary, in this Letter we have presented the first systematic numerical study of the effect of impurities on bulk turbulence in W7-X and, for comparison, in LHD and ITER. By means of nonlinear gyrokinetic simulations, we have demonstrated that impurities can effectively alter turbulent heat transport. This finding is robust and manifests both for electron and, more substantially, ion heat fluxes in the three configurations. This has obvious interest for future reactors, as the main ions are the reactants in the fusion fuel.
When the density gradient of impurities and main species are, roughly, parallel (resp. anti-parallel), turbulent fluctuations and ion heat losses decrease (resp. increase). As for the electron heat flux, the combination of the impurity effect and the response of the configuration to modifications of $a/L_{n_e}$ forced by quasineutrality, introduces differences across the devices, which will be thoroughly addressed in upcoming studies.
A central message of the present work is that, regardless of the magnetic configuration, the ion heat flux reaches a minimum value at a specific impurity concentration, that could be seen as an optimal impurity content to operate with as long as radiative losses were tolerable. 
As a particularly interesting case, in W7-X, when $Q_i$ is reduced, $Q_e$ is also reduced and the impurity heat flux remains low.
In summary, the present study emphasizes the ambivalent role of impurities and their potential to enhance or degrade the performance of magnetically confined plasmas through its active role on turbulence.

J.M.G.R. is grateful to Michael Barnes for his support with the code \texttt{stella} and to Arturo Alonso and Felix Reimold for fruitful discussions. This work has been carried out within the framework of the EUROfusion Consortium, funded by the European Union via the Euratom Research and Training Programme (Grant Agreement No 101052200 – EUROfusion). Views and opinions expressed are however those of the author(s) only and do not necessarily reflect those of the European Union or the European Commission. Neither the European Union nor the European Commission can be held responsible for them. This research was supported in part by grant PID2021-123175NB-I00, Ministerio de Ciencia e Innovación, Spain. This work was supported by the U.S. Department of Energy under contract number DE-AC02-09CH11466. The U.S.~Government retains a non-exclusive, paid-up, irrevocable, world-wide license to publish or reproduce the published form of this manuscript, or allow others to do so, for U.S.~Government purposes. Simulations were performed in the supercomputers Marconi (CINECA, Italy) and Marenostrum (Barcelona Supercomputing Center, Spain).
%%\bibliographystyle{plainnat}
%\bibliography{Bibliography}% Produces the bibliography via BibTeX.

\section{Appendix: magnetic configurations, plasma parameters, numerical convergence and normalization convention}
\label{sec:supplemental}

\begin{figure*}[ht]
	\includegraphics[width=0.32\textwidth]{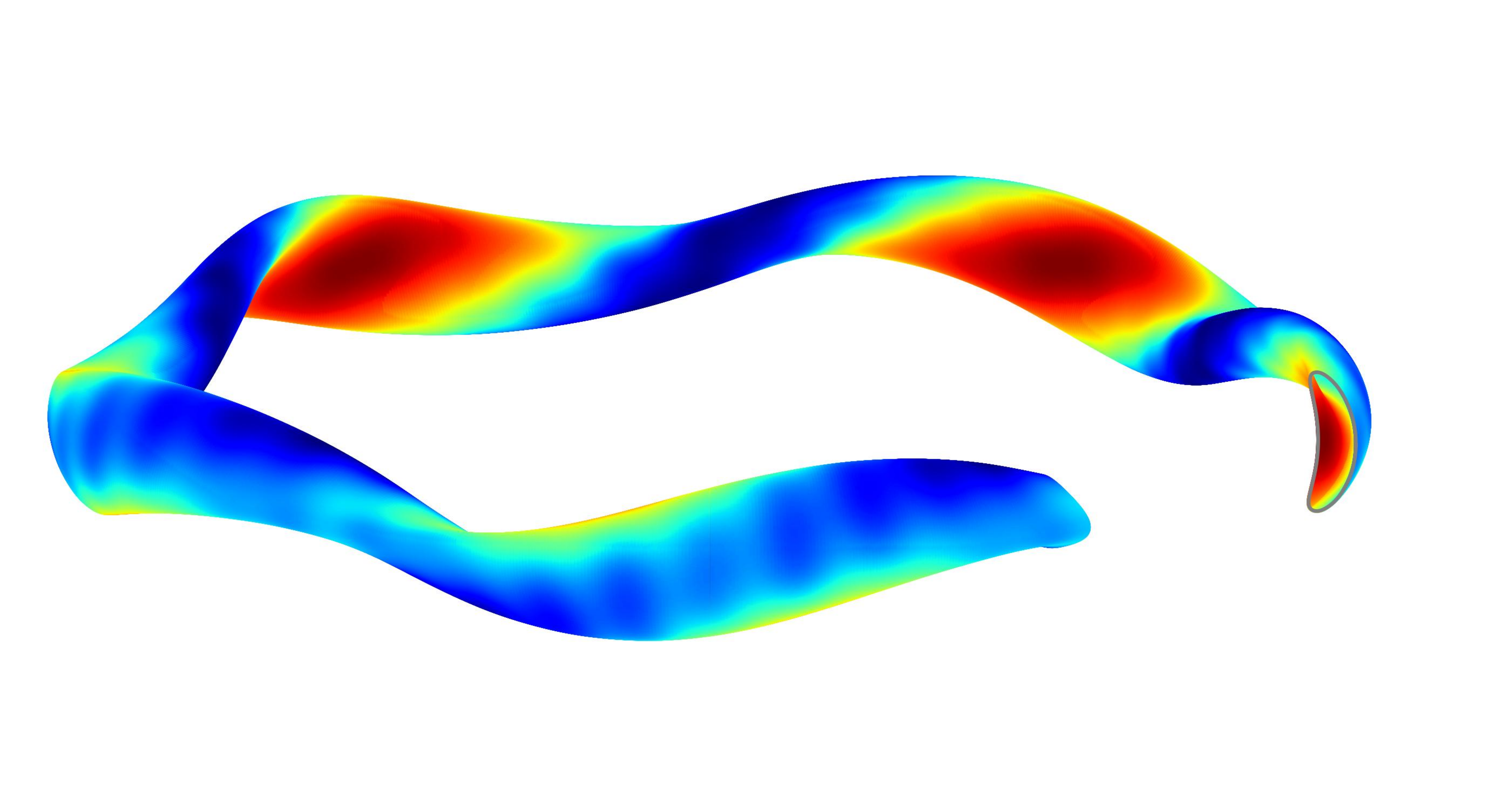}
	\includegraphics[width=0.32\textwidth]{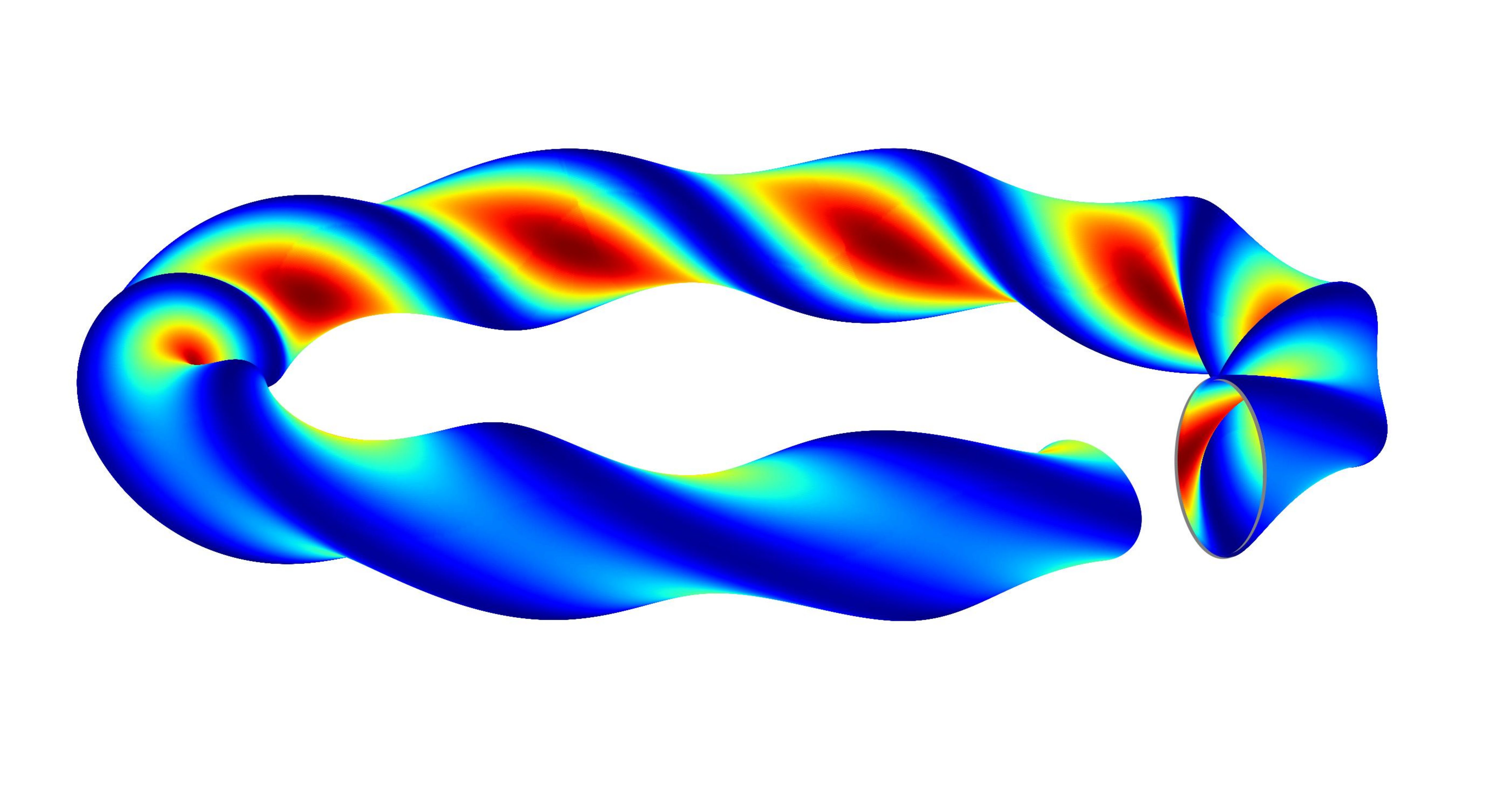}
	\includegraphics[width=0.32\textwidth]{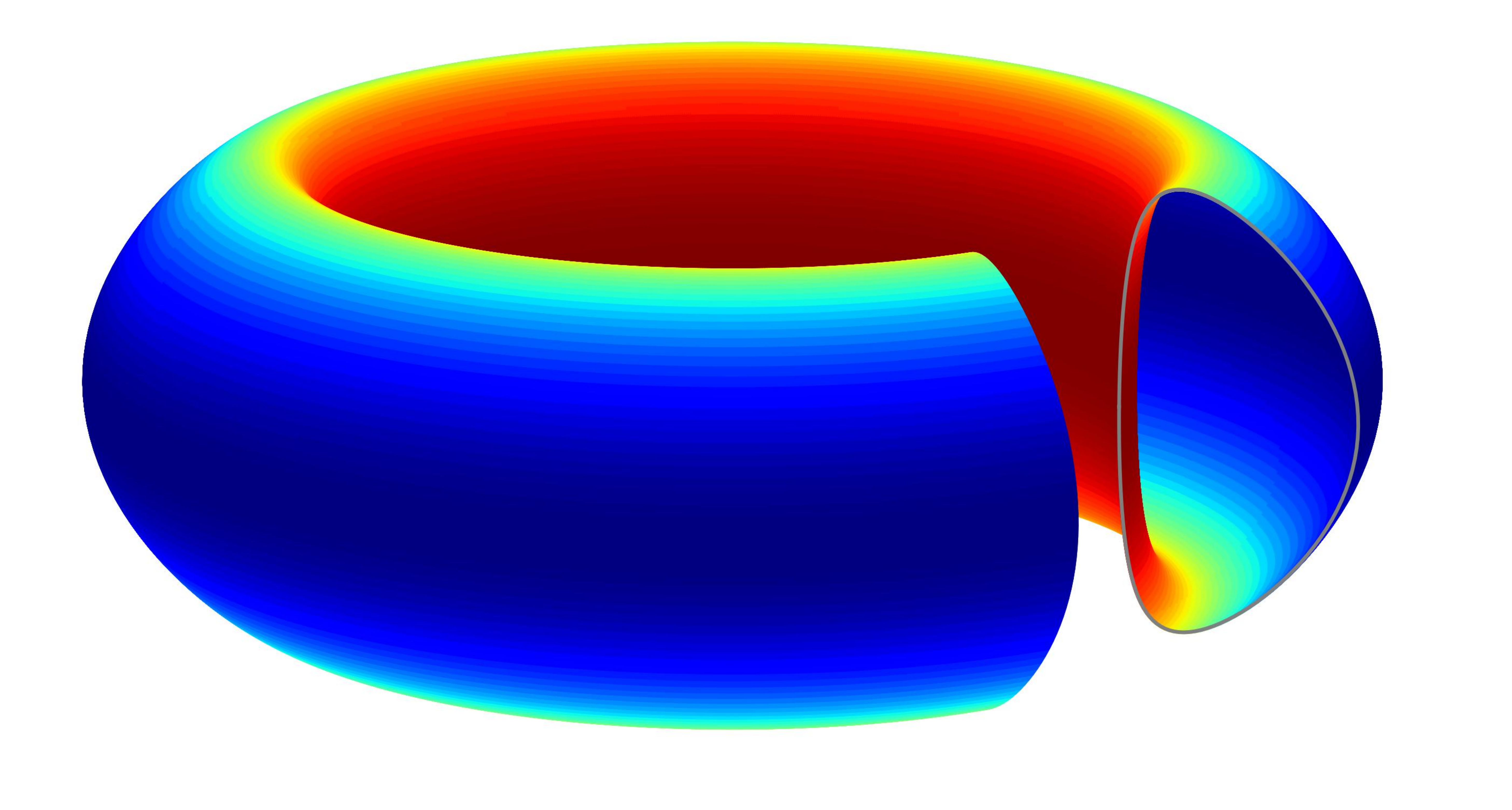}
	\caption{\label{fig:configurations} 
		Magnetic field strength---minimum is blue, maximum is red---on the flux surface r/a = 0.7 for W7-X (left), LHD (center) and ITER (right).}
\end{figure*}

\begin{figure*}
	\includegraphics{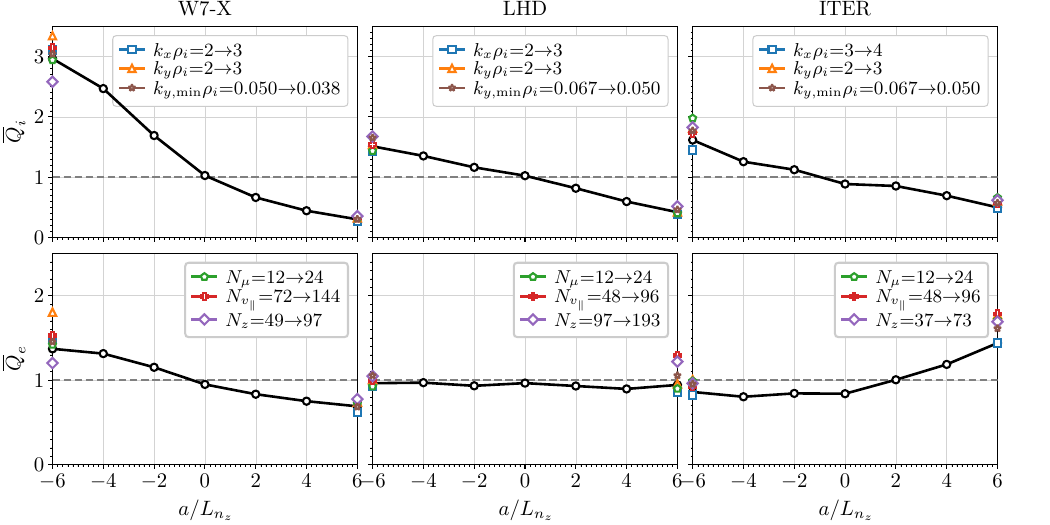}
	\caption{\label{fig:resolution} 
		The results already shown in Fig.~2(a) and Fig.~2(b) for, respectively, the ion (top) and electron (bottom) heat fluxes for W7-X (left), LHD (center), and ITER (right), are depicted by solid black lines. For the two extreme cases, $a/L_{n_z}=-6$ and $a/L_{n_z}=6$, the points represent differences resulting from increases in simulation resolution compared to the set of parameters considered in the Letter for each specific device.}
\end{figure*}

Three magnetic configurations have been considered in the present work: the standard configuration of \mbox{W7-X} (also known as the EIM configuration), with minor radius $a=0.513$ m, major radius $R_0=5.52$ m, rotational transform $\iota=0.89$ and global magnetic shear $\hat{s}=-0.13$; one of the the so-called \textit{inward shifted} LHD configurations, specifically, that with $\left\{a, R_0, \iota, \hat{s}\right\}=\left\{0.64, 3.67, -1.45, -1.33\right\}$; and a tokamak equilibrium with similar geometric parameters to those foreseen and modeled for ITER scenarios \cite{Kessel2007, Campbell2016}, in particular, with  $\left\{R_0/a, q, \hat{s}, \kappa, \delta\right\}=\left\{3, 2.2, 2.3, 1.85, 0.5\right\}$, where $\kappa$ and $\delta$ are the plasma elongation and triangularity, respectively, and $q=1/\iota$ is the safety factor. Here, all parameters, except $a$ and $R_0$, refer to their values at the simulated radial position $r/a=0.7$. Whereas the stellarator equilibria are passed to \texttt{stella} through a global equilibrium output generated by the \texttt{VMEC} code \cite{Hirshman1983}, the ITER configuration is defined through a set of Miller parameters for modeling local tokamak equilibria \cite{Miller1998}. In all three cases, the simulated flux tube corresponds to that extending symmetrically with respect to the equatorial plane (with poloidal angle $\theta=0$), and the toroidal plane $\zeta=0$, with $\zeta$ the toroidal azimuth. The flux tube has been extended for one poloidal turn.

The selected radial position, $r/a=0.7$, is a representative position for the inner plasma edge, where the largest turbulent fluctuations and heat losses are typically measured \cite{Baehner2021, Estrada2021, Bozhenkov_nf_2020} and numerically predicted \cite{BanonNavarro2020, GonzalezJerez2024} in W7-X.  In that region, the temperature gradient of the bulk species tends to be appreciably larger than the density gradient, except in scenarios with strong external particle fueling, where they equilibrate \cite{Carralero2021}. Noticeable temperature gradients and weakly peaked density profiles are common in LHD as well (see e.g. \cite{Takahashi2018}) and are foreseen in ITER \cite{Kessel2007}. Hence, the choice $a/L_{n_i}=1$ and $a/L_{T_i}=a/L_{T_e}=3$ serve as both a sensible case study and as a basis for comparison between the three devices.

Regarding the impurity parameters, thermalization of impurities and main ions is assumed, thus $a/L_{T_z}=3$ has also been fixed in all simulations. With regard to the impurity density profile, its characterization is far less routine and accurate compared to the main plasma parameters. Volume-averaged values of $Z_{\mathrm{eff}}$ are regularly measured in most present-day devices (the value $Z_{\mathrm{eff}}=1.4$ for the density gradient scans is based on the typical value measured during the first campaigns of \mbox{W7-X}, with carbon being the main intrinsic impurity \cite{Beidler_nature_2021}, although values above $Z_{\mathrm{eff}}=3.0$ have been measured in experiments with deliberate boron injection \cite{Lunsford_2021}). However, the specific contribution of each impurity to the effective charge and its radial distribution is limited to some selected elements and charge states. These uncertainties and the fact that impurity behaviour can strongly differ depending on the scenario and plasma conditions have motivated the broad scans on the impurity density gradient and concentration in this work. As a paradigmatic case, scenarios leading to both strong impurity  accumulation with highly peaked impurity density profiles and strongly hollow impurity profiles have been reported in LHD \cite{Nakamura2017}. With hollow impurity density profiles being a rarity so far, flat \cite{Langenberg2021, Romba2023} and strongly peaked impurity profiles have been the most commonly observed situations in W7-X.

Finally, the resolution of the simulations has been considered such that increasing it substantially through each relevant parameter one by one, the results changed quantitatively by a small percentage. This criterion has been checked for the two extreme cases of the impurity density gradient scan shown in Fig.~2(a) and Fig.~2(b).
Before presenting the results from these checks, it is worth providing a concise overview of the coordinates employed by \texttt{stella}.
The coordinates $x$ and $y$ measure distance from the center of the flux tube in the radial and binormal directions, respectively, and are treated by the code in Fourier space, with $k_x$ and $k_y$ being the associated wavenumbers. The parallel coordinate that measures distance along the magnetic field line is generically denoted by $z$, and can be chosen to be the toroidal azimuth, the arc-length along the magnetic field line, or a poloidal coordinate. Regarding velocity coordinates, $v_{\|}$ is the velocity component parallel to the magnetic field, and $\mu$ is the magnetic moment. 
Fig.~\ref{fig:resolution} depicts the change in the ion and electron heat fluxes when, for each specific equilibrium, the largest radial and binormal wavenumbers, $k_{x,\mathrm{max}}$ or $k_{y,\mathrm{max}}$, are increased; or the step in $k_y$ ($\Delta k_{y}$) is decreased (the step in $k_x$, due to the application of the standard twist and shift parallel boundary conditions, decreases accordingly); or the number of grid points in $z$, $\mu$ and $v_{\|}$ are also increased. As can be observed, such resolution enhancements neither yield qualitative changes nor alter the results quantitatively by a large amount. The electron heat flux appears to be more sensitive to the changes though, particularly to the increase of the $k_y$ domain. This is not surprising, given that in this work we are simulating ion Larmor scale turbulence, omitting the contribution to the electron heat flux at electron Larmor scales, toward which the electron heat flux spectrum tends to ramp up already in the simulated range. 

Note that wavenumbers are normalized by the inverse of the thermal ion Larmor radius, $\rho_i=v_{\mathrm{th},i}/\Omega_i$,  with $v_{\mathrm{th},i}=\sqrt{2T_i/m_i}$ the thermal speed of the main ions, $\Omega_i=e B_r/m_i$ their gyrofrequency, $e$ the unit charge, $m_i$ the ion mass, $T_i$ the ion temperature and $B_r$ a reference magnetic field. 
In the Letter, heat fluxes are normalized to the gyro-Bohm value for the main ions, $Q_{i,\mathrm{gB}}=(\rho_i/a)^2 n_i T_i v_{\mathrm{th},i}$. For further details on the equations solved by \texttt{stella}, coordinates used, geometry treatment and normalization conventions, see \cite{Barnes_jcp_2019, GonzalezJerez_jpp_2022}.

\end{document}